\documentclass[conference]{IEEEtran}
\IEEEoverridecommandlockouts

\usepackage{cite}
\usepackage{amsmath,amssymb,amsfonts}
\usepackage{algorithmic}
\usepackage{graphicx}
\usepackage{textcomp}
\usepackage{xcolor}
\usepackage{tabularx} 

\usepackage{mdwlist}
\usepackage[ruled,vlined,shortend, linesnumbered]{algorithm2e} 
\usepackage[]{algorithmic} 
\usepackage{booktabs}
\usepackage{color}

\usepackage{amsthm,amssymb}
\usepackage{amsmath}
\usepackage[font=small]{caption}
\usepackage[skip=3pt]{subcaption}
\usepackage{multirow}
\usepackage{mathrsfs}
\usepackage{amsbsy}
\usepackage[mathscr]{eucal} 
\usepackage[flushleft]{threeparttable}
\usepackage[hidelinks]{hyperref}
\usepackage{comment}
\usepackage{paralist}
\usepackage{pifont}
\usepackage{bbding} 
\usepackage{balance}

\usepackage{bm}
\definecolor{myblue}{HTML}{ECECFF}
\definecolor{myred}{HTML}{FFE1E1}

\usepackage{thmbox} 
\newtheorem[M, bodystyle=\normalfont\noindent]{problem}{Problem} 

\usepackage{tikz}
\usetikzlibrary{arrows.meta}

\def\BibTeX{{\rm B\kern-.05em{\sc i\kern-.025em b}\kern-.08em
    T\kern-.1667em\lower.7ex\hbox{E}\kern-.125emX}}

\newcommand{\hide}[1]{}

\newcommand{\bit}{\begin{compactitem}}
\newcommand{\eit}{\end{compactitem}}
\newcommand{\ben}{\begin{compactenum}}
\newcommand{\een}{\end{compactenum}}

\usepackage{enumitem}
\usepackage[protrusion=true,expansion=true]{microtype}
\pdfoutput=1

\title{Voltage-Regulated Sparse Optimization for Proactive Diagnosis of Voltage Collapses
}

\author{Qinghua Ma$^1$, Seyyedali Hosseinalipour$^1$, Ming Shi$^1$, Jan Drgona$^2$, and Shimiao~Li$^1$\\
	$^1$University at Buffalo--SUNY, NY, USA\\$^2$Johns Hopkins University, MD, USA
    \thanks{\noindent Manuscript submitted to 2026 IEEE PES General Meeting.} 
}

\begin{document}
\maketitle

\begin{abstract}
	This paper aims to proactively diagnose and manage the voltage collapse risks, i.e.,  the risk of bus voltages violating the safe operational bounds, which can be caused by extreme events and contingencies. We jointly answer two resilience-related research questions: \textit{(Q1) Survivability:} Upon having an extreme event/contingency, will the system remain feasible with voltage staying within a (preferred) safe range? 
\textit{(Q2) Dominant Vulnerability:} If voltage collapses, what are the dominant sources of system vulnerabilities responsible for the failure? This highlights some key locations worth paying attention to in the planning or decision-making process. To address these questions, we propose a voltage-regulated sparse optimization that finds a minimal set of bus locations along with quantified compensations (corrective actions) that can simultaneously enforce AC network balance and voltage bounds. Results on transmission systems of varying sizes (30-bus to 2383-bus) demonstrate that the proposed method effectively mitigates voltage collapses by compensating at only a few strategically identified nodes, while scaling efficiently to large systems, taking on average less than 4 min for 2000+ bus cases. This work can further serve as a backbone for more comprehensive and actionable decision-making, such as reactive power planning to fix voltage issues.
\end{abstract}


\section{Introduction}
\label{sec:Introduction}
\begin{figure*}[h]
\centering
\subfloat[]
{%
  \includegraphics[clip,width=0.33\linewidth]{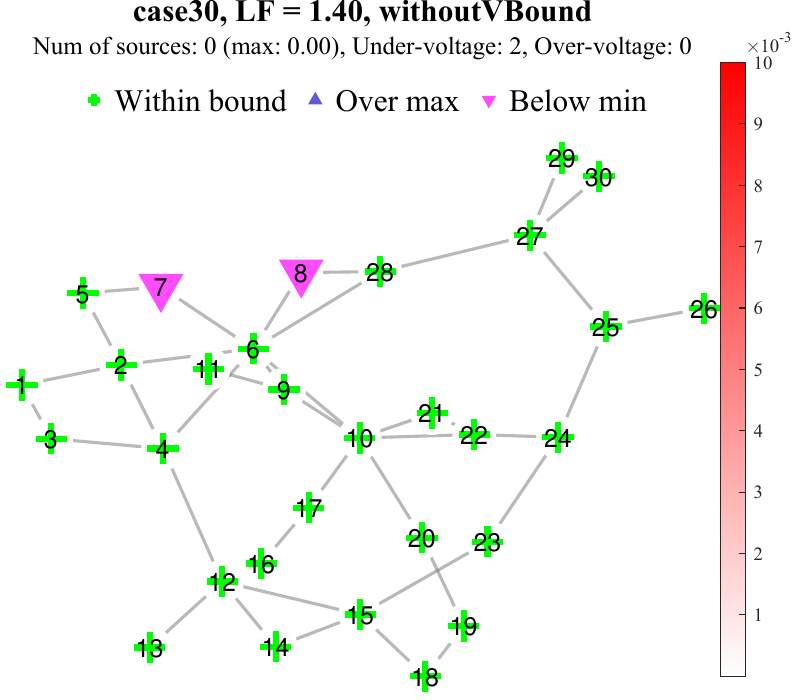}%
}\hfill
\subfloat[]{%
  \includegraphics[clip,width=0.33\linewidth]{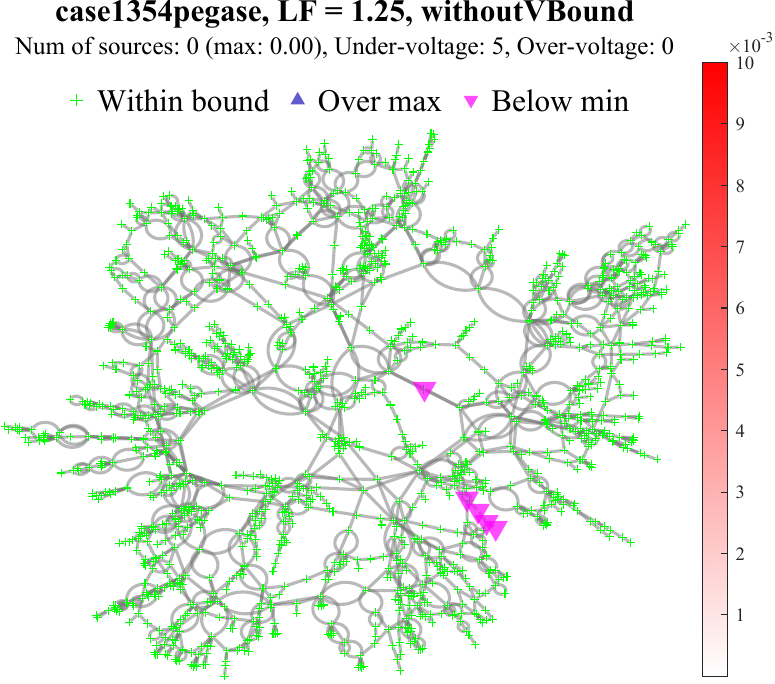}%
}\hfill
\subfloat[]{%
  \includegraphics[clip,width=0.33\linewidth]{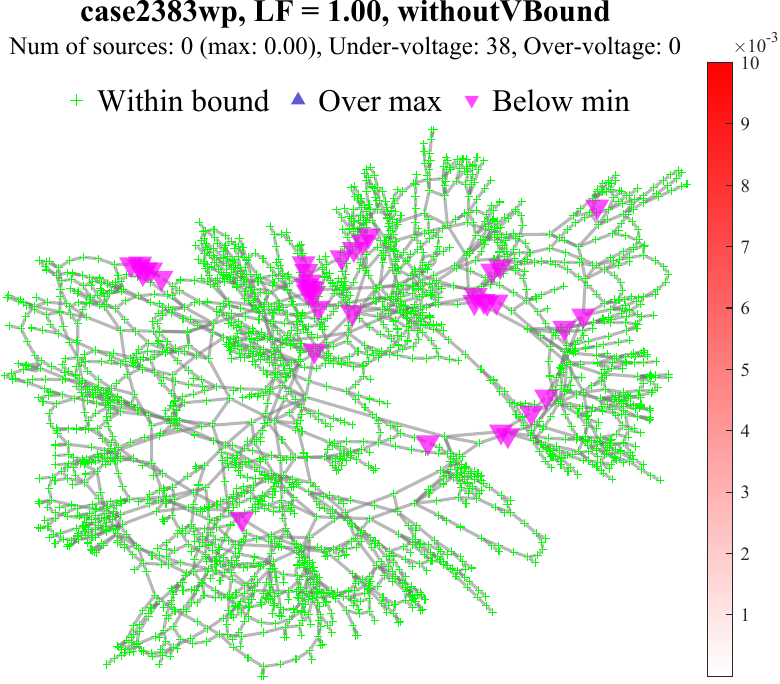} \label{fig: sparse graph + VBound c}%
}\\
\vspace{5pt}

\subfloat[]{%
  \includegraphics[clip,width=0.33\linewidth]{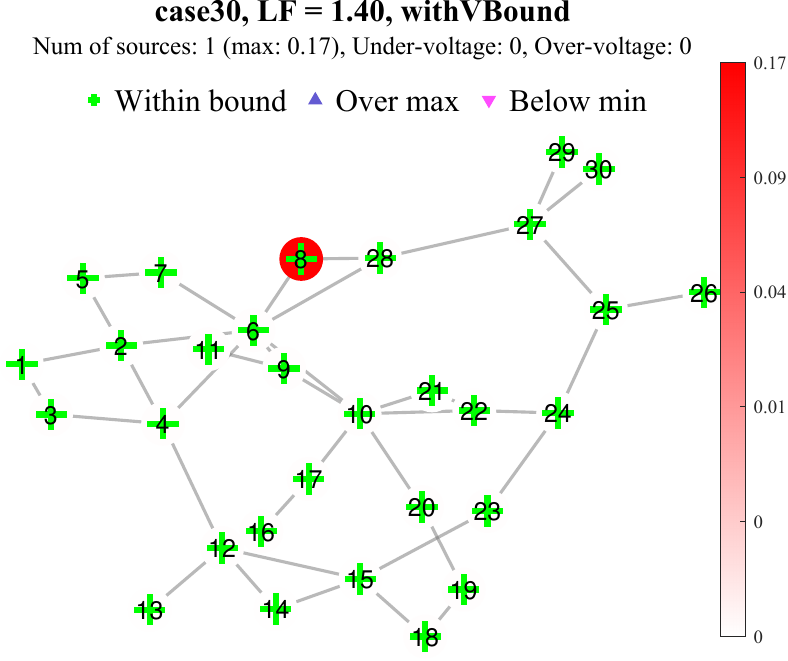}%
}\hfill
\subfloat[]{%
  \includegraphics[clip,width=0.33\linewidth]{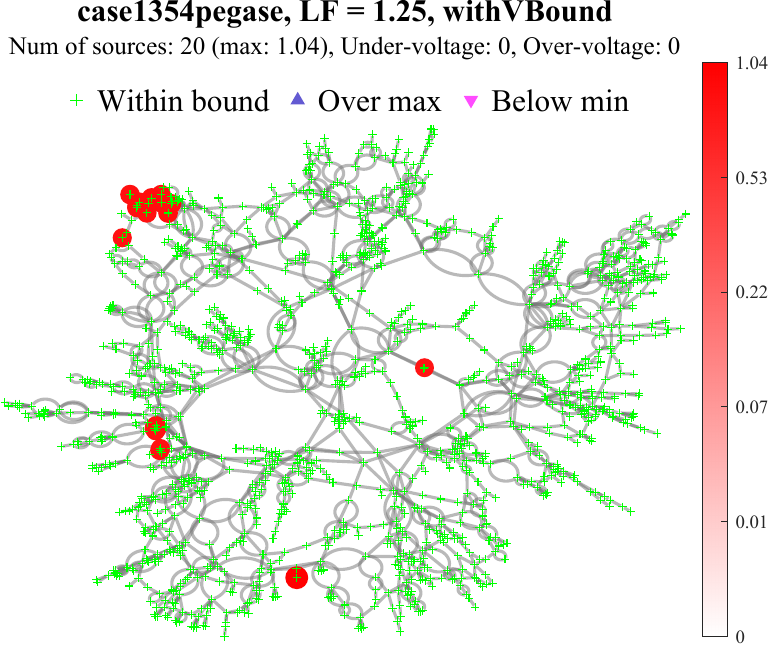}%
}\hfill
\subfloat[]{%
  \includegraphics[clip,width=0.33\linewidth]{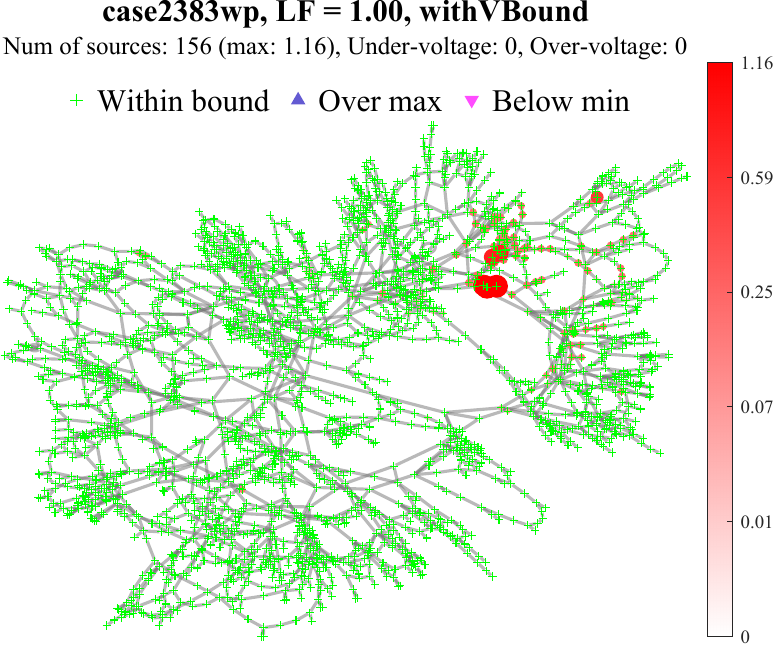} \label{fig: sparse graph + VBound f}%
}
\caption{{Voltage collapse and restoration.} \textit{Top Row (baseline simulation)}: Voltage collapses occur under high demand conditions. \textit{Bottom Row (proposed method)}: The proposed voltage-regulated sparse optimization identifies a few key system vulnerabilities; and compensations at a few identified buses can fix the issue. Red markers denote the magnitude of compensation sources (in per-unit current) needed at a few identified nodes to keep buses voltage within the predefined safe range, localizing and quantifying instability vulnerabilities. In Fig. \ref{fig: infeasible, sparse graph + VBound}, we further illustrate how both power imbalance and voltage collapse can be simultaneously fixed.}
\vspace*{-1em}
\label{fig: sparse graph + VBound}
\end{figure*}

Electrical power grids are increasingly challenged by extreme events, equipment failures, and cyberattacks\cite{grid-security-Vyas}. As a result, these systems can suffer from major issues ranging from violations of operational bounds (e.g, voltage bounds and transmission line limits) to system collapses (e.g., power outage or blackouts). To understand and anticipate such events, operators and researchers often rely on simulation tools that can model and analyze grid behavior under various stress conditions. Among these, traditional power flow simulators \cite{powerflow-diverge} are of utmost importance. These simulators rely on solving the AC network balance equations $\bm{g}(\bm{v})=\bm{0}$ for feasible bus voltages $\bm{v}$ using the Newton-Raphson method, and are routinely employed to perform \textit{what-if} analyses and identify problematic regions in the grid, such as under-voltage buses or overloaded transmission lines. Nevertheless, during system collapses or blackouts, the solver typically diverges due to the absence of feasible solutions. 

To overcome this limitation, prior works \cite{sugar-pf,sugar-pf-Amrit,infeasibility-quantified-pf-trad} proposed \textit{infeasibility-quantified power flow} methods, which introduce slack variables to quantify the deficiency of power supply and allow convergence on otherwise collapsed (blackout) systems. The resulting solution of these methods provides a dense vector of slack variables to quantify the power deficiency at each bus, offering a measure of blackout severity. Although these methods marked an important advancement in power grid analysis, they provide little insight into the \textit{locations of the dominant source of failure or system vulnerabilities} responsible for system collapses and how to fix them. To offer targeted insights, later work \cite{SparseFeas} proposed a sparse optimization that efficiently identifies a minimal set of critical locations along with quantified compensations to restore power balance, thereby locating a few key system vulnerabilities. The sparse diagnosis framework has been extended to distribution networks \cite{SparseFeas4DNet}, combined transmission-distribution networks \cite{SparseFeas4TnDNet}, and multi-period growing load stress settings \cite{sparsefeas-mp}. 

While these extensions significantly broadened the applicability of sparse optimization in identifying critical vulnerabilities, the results may not always translate into practically actionable mitigation strategies. This is because they primarily focused on recovering from blackout collapses by enforcing the AC power balance as equality constraints, without explicitly considering other operational limits such as maintaining voltage within safe bounds. In reality, maintaining a healthy voltage profile is of vital importance to power grid reliability, as violations like under-voltage can easily overheat electronic devices to cause equipment damages, instability, and even trigger cascading outages. Under-voltage problems often arise under high real or reactive power demand conditions, which lead to increased line currents during power delivery, and, in turn, cause larger voltage drops across lines, ultimately reducing the voltage at load locations. This relationship is accurately characterized by power system P-V and Q-V curves, which trace voltage versus real and reactive power at a bus. While a previous work\cite{DenseFeas-VBound} has introduced voltage bounds into infeasibility-quantified power flow \cite{sugar-pf} to identify the minimal reactive power compensation needed to resolve voltage violations, it formulated a dense (instead of sparse) optimization, producing corrective actions spread across the entire network.

This paper bridges the gap by jointly addressing  the following resilience questions:
\begin{itemize}[leftmargin=4mm]
    \item \textit{\textbf{(Q1) Survivability:}} Given an extreme event/contingency, will the power system remain feasible with voltage staying within a (preferred) safe range? 
    \item \textit{\textbf{(Q2) Dominant Vulnerability:}} If voltage collapses, where are the dominant sources of vulnerabilities responsible for the failure? Equivalently, how to apply compensations (corrective actions) at only a few identified nodes to fix it (i.e., restoring normal operations)?
\end{itemize}
Insights from (Q2) are highly valuable, as accurate and efficient localization of such dominant set of nodes pinpoints the deficiency of power and
highlights some critical locations for special attention in the
planning process. For instance, consider reactive power
planning (RPP) problems \cite{sparsePFref5-planningFACTS4congestion,sparsePFref7-planningSVC,sparsePFref8-planningFACTS,sparsePFref6-planningSVCnTCSC} that aim to find the optimal
allocation of reactive power support through capacitor banks or FACTS devices such as static VAR compensators (SVC).
Such problems correspond to finding the sparsest reactive
power compensation vector that satisfies system power
balance and operation limits in an optimization-based power
flow study. 

To this end, we propose, for the first time in the literature, a voltage-regulated sparse optimization framework that incorporates voltage operational bounds into the sparse diagnosis formulation. The method minimizes slack compensation sources while enforcing sparsity, subject to both AC network constraints and voltage limits, and leverages circuit-theoretic modeling and optimization heuristics to ensure scalability. The framework can be further extended to include other operational limits and controllable resources.

We evaluate the efficacy and scalability of our proposed method on IEEE standard transmission systems under voltage collapses induced by high load demand. Key results in Fig. \ref{fig: sparse graph + VBound} demonstrate that localized compensation at a small subset of buses can ensure bounded voltage for the entire system. For instance, Fig. \ref{fig: sparse graph + VBound}(b) and Fig. \ref{fig: sparse graph + VBound}(e) illustrate that, under a 25\% increase in demand, the 1354-bus system exhibits under-voltage at five buses, and stability can be restored through compensations applied at merely 20 of all 1354 buses.

\section{Background}
\label{sec:bkg}
\subsection{Circuit-Theoretic Modeling of Power System}\label{bkg: ckt formulation}

Recent advances have introduced circuit-theoretic modeling for power-flow and optimization problems, inspired by classical circuit simulators such as SPICE \cite{sparsePFref15-spice}. Unlike conventional power-based models that use state variables $(P,Q,V)$ or variables $(|V|,\delta)$ in the polar coordinates, circuit-theoretic modeling represents each power component based on its current-voltage (I-V) relationship in the Cartesian coordinates with state variables $(V^\mathrm{Real},V^\mathrm{Imag})$. This formulation yields linear models for shunt, transmission line, transformer, and slack bus, while producing nonlinear models for PV generators and PQ loads. Subsequently, for each bus $i$, by combining I-V formulations of various components according to the system structure, network balance equations with respect to bus voltage $\bm{v}$, i.e., $\bm{g}(\bm{v})=\bm{0}$, can be obtained based on the Kirchhoff's Current Law (KCL). It is worth noting that such circuit-theoretic formulations apply uniformly to transmission and distribution systems and have been successfully applied to power flow \cite{sugar-pf}, state estimation \cite{SUGAR-SE-Li, convexSE-LAV-Li, ckt-GSE,BayesGSE}, and optimal power flow \cite{sparsePFref12-SUGAR-opf}. This work likewise adopts the circuit-theoretic steady-state modeling and applies circuit-inspired optimization heuristics\cite{sugar-pf-Amrit}.

\subsection{Infeasibility Quantified for Collapsed Systems}
Previous works \cite{sugar-pf,sugar-pf-Amrit} introduced a dense optimization model (\textit{Problem \ref{prob: dense optimization}}) that converges even for a collapsed system by minimizing slack sources $\bm{n}=[n_i]$ which can restore AC network balance $\bm{g}(\cdot)$. Under the circuit formulation, $\bm{v}=[V_i]$ consists of voltages $V_i=V_i^{\text{Real}}+\mathrm{j}V_i^{\text{Imag}}, \forall \text{ bus } i$; and slack source $n_i=n_i^{\text{Real}}+\mathrm{j}n_i^{\text{Imag}}$ denotes current injection at $\text{bus } i$. 
\begin{problem}[Dense Optimization] \label{prob: dense optimization} 
    \begin{equation}
    \min_{\bm{v},\bm{n}}~ \frac{1}{2} ||\bm{n}||_{_2}^2,~~~ \text{ s.t. }~ g_i(\bm{v}) + n_i = 0,\ \forall \text{ bus }i\label{eq: dense optimization}
\end{equation}
\end{problem}

\subsection{Sparse Diagnosis for Collapsed Systems}
To localize system vulnerabilities, prior work \cite{SparseFeas} extended \textit{Problem~\ref{prob: dense optimization}} to a sparse framework which iteratively solves \textit{Problem~\ref{prob: sparse diagnosis}} below with updated sparsity coefficients $c_i, \forall i$, until reaching a highly sparse compensation vector $\bm{n}$ to restore balance for a collapsed system.
\begin{problem} [Sparse Optimization] \label{prob: sparse diagnosis}
\begin{equation}
    \min_{\bm{v},\bm{n}}~ \frac{1}{2}||\bm{n}||_{_2}^2 + \sum_i c_i|n_i|,~~\text{ s.t. }~ g_i(\bm{v}) + n_i = 0,\ \forall\text{ bus } i \label{eq: sparse optimization}
\end{equation}
\end{problem}

Prior work\cite{SparseFeas} demonstrated sparse diagnosis concept in large systems up to the size of the Eastern Interconnection network ($>$80,000 buses) and recognized only one dominant failure source (vulnerability) when the load demand increased by $7\%$. But this formulation only considers power balance but not voltage violations. This work aims to address this issue. 

\section{Method}
\label{sec:method}
To identify dominant feasibility sources responsible for both system collapse and voltage violations, we aim to integrate voltage bound $V_{i\text{, min}}\leq|V_i|\leq V_{i\text{, max}}$ for each bus $i$ into the sparse \textit{Problem~\ref{prob: sparse diagnosis}}. Under a circuit-theoretic formulation, $|V_i|=\sqrt{(V_i^\text{Real})^2+(V_i^\text{Imag})^2}$.  We introduce an intermediate variable $V_{i\text{, sq}}$ to denote square of voltage magnitude ($V_{i\text{, sq}}=|V_i|^2$) at each bus. With preliminaries defined as above, we propose the following \textit{voltage-regulated sparse optimization}:
\begin{problem} [Voltage-Regulated Sparse Optimization] \label{prob: voltage-regulated sparse optimization}
\begin{subequations}\label{eq: voltage-regulated sparse optimization}
\vspace{-3mm}
    \begin{align}
        & \min_{\bm{v},\bm{n}, \bm{V}_\text{sq}}~ \frac{1}{2}||\bm{n}||_{_2}^2 + \sum_i c_i|n_i| \label{eq: voltage-regulated sparse optimization a}\\
        \text{s.t. }~~& g_i(\bm{v}) + n_i = 0,\ \forall i \label{eq: voltage-regulated sparse optimization b}\\
        & V_{i\text{, sq}} = (V_i^\mathrm{Real})^2+(V_i^\mathrm{Imag})^2,\ \forall i \label{eq: voltage-regulated sparse optimization c}\\
        &V_{i\text{, min}}^2\leq V_{i\text{, sq}} \leq V_{i\text{, max}}^2,\ \forall i \label{eq: voltage-regulated sparse optimization e}
    \end{align}
\end{subequations}
\end{problem}

Solving this problem at large scale incurs several key challenges: (i) non-differentiability induced by the L1-norm based sparsity term $\sum_i c_i|n_i|$; (ii) nonlinearity and nonconvexity induced by the nonlinear constraints (3b)-(3c); and (iii) the need for promoting \textit{highly sparse} solutions  $\bm{n}$ subject to the aforementioned constraints. 

To handle (i), i.e., the non-differentiability of the objective function \eqref{eq: voltage-regulated sparse optimization a}, we introduce an extra slack vector $\bm{t}=[t_i]$ and transform the above problem to an equivalent differentiable form: $\min \frac{1}{2}||\bm{n}||_{_2}^2 + \sum_i c_i{t}_i, \text{ s.t. }(\ref{eq: voltage-regulated sparse optimization b})-(\ref{eq: voltage-regulated sparse optimization e})$, $-\bm{t} \preceq \bm{n}\preceq \bm{t}$.

Further, to address (ii), i.e., facilitating the convergence of the nonlinear nonconvex problem, we implement a circuit-based interior point method \cite{sugar-pf-Amrit} along with the newly added voltage-related constraints. A set of perturbed Karush-Kuhn-Tucker (KKT) conditions are formed and solved using the Newton-Raphson method with circuit-inspired optimization heuristics (including voltage limiting and damping) \cite{sugar-pf-Amrit} adopted at each Newton-Raphson iteration.

Finally, to tackle (iii), i.e., facilitating a highly-sparse solution for large systems, we split the original problem into a series of subproblems where each sub-problem easily converges within a small number of iterations and reaches a higher sparsity level than the previous sub-problem.  In particular, the subproblem series begins with a dense optimization $\min_{\bm{v},\bm{n},\bm{V}_\text{sq}}\frac12||\bm{n}||_{_2}^2$ to obtain a strictly feasible initial point for the subsequent sparse subproblems. Each sparse subproblem solves \textit{Problem \ref{prob: voltage-regulated sparse optimization}} with updated sparsity coefficients and uses the previous solution as its initial point. When updating sparsity coefficients for each subproblem, we leverage the efficient sparsity-enforcing mechanism proposed in \cite{SparseFeas}, which adaptively toggles the location-wise sparsity coefficient $c_i$ for each bus $i$ between a relatively larger value $c_\mathrm{H} = 10$ and a smaller value $c_\mathrm{L} = 0.1$, according to the most updated vector $\bm{n}$. This adaptive mechanism creates uneven penalties across locations and promotes highly sparse yet numerically stable solutions. Details on the selection and theoretical justification of $c_\mathrm{H}$ and $c_\mathrm{L}$ from an optimization standpoint can be found in~\cite{SparseFeas}.

\section{Experiments}
\label{sec:experiment}
We test and experiment on standard IEEE transmission systems: \textsc{case30}, \textsc{case1354pegase} and \textsc{case2383wp}. For each test case, we create voltage collapse conditions by increasing the load demand (via load factor). We compare two methods:
\begin{enumerate}[leftmargin=4.8mm]
    \item \textbf{Without VBound (baseline)}: This method uses and solves {sparse optimization} \textit{Problem \ref{prob: sparse diagnosis}}.
    \item \textbf{With VBound (proposed)}: This method uses and solves voltage-regulated sparse optimization \textit{Problem \ref{eq: voltage-regulated sparse optimization}}.
\end{enumerate}

Notably, although this paper mainly experiments on uniformly increased load demand, the proposed work itself is agnostic to event or demand patterns and applies generally to any extreme event or disturbance (e.g., a generator outage or sudden load increase at some locations) that can result in voltage issues. All experiments were conducted on a Microsoft Windows 11 laptop with 2.70 GHz Intel Core Ultra 9 275HX and 32 GB RAM.

\subsection{Voltage Regulation Effect}
\begin{figure}[h]
    \centering
    \includegraphics[width=\linewidth]{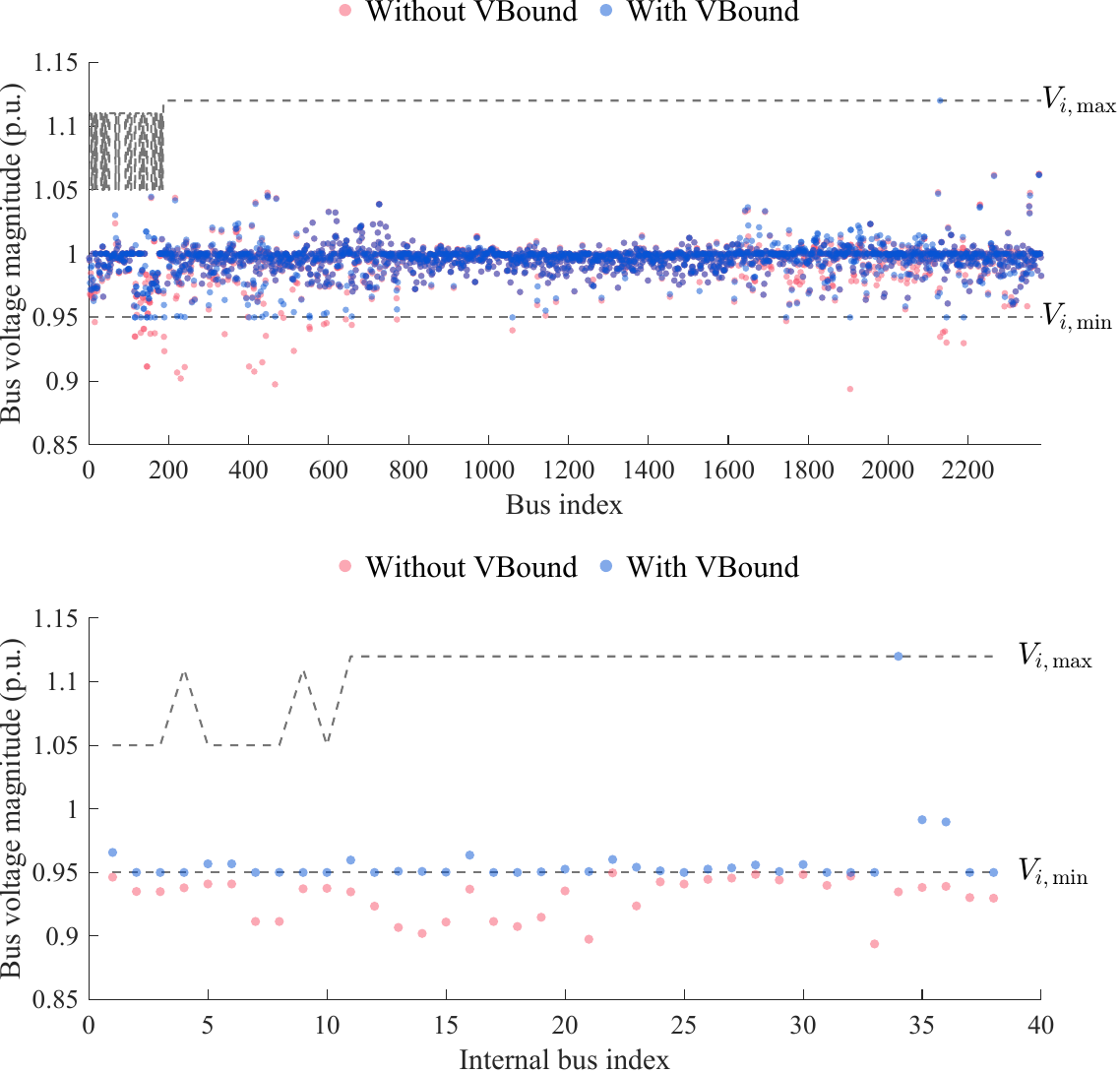}
    \caption{Bus voltage magnitudes from the baseline and proposed methods for \textsc{Case2383wp}. The \textit{Top} plot shows all buses, while the \textit{Bottom} plot specifically focuses on the voltage-regulated buses (bus has voltage violations that are fixed by the proposed approach). The proposed method restores all violated voltages within a safe range: some exactly at the lower bound and others strictly within bounds.}
    \label{fig: voltage change}
\end{figure}

\begin{figure}[h]
    \centering
    \includegraphics[width=\linewidth]{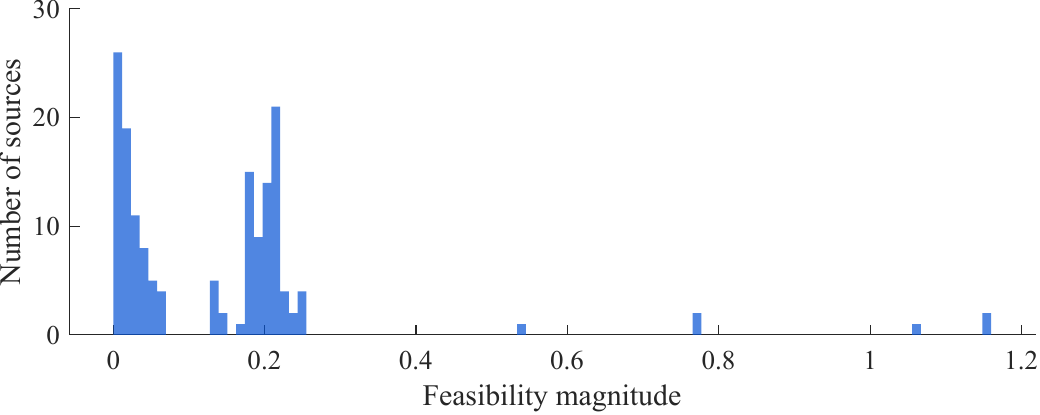}
    \caption{Histogram of the 200 slack sources $\bm{n}$ for \textsc{case2383wp} (in Fig. \ref{fig: sparse graph + VBound f}): only a few buses require significant adjustments.}
    \label{fig: feasibility histogram}
\end{figure}

We first evaluate the voltage regulation capability of our proposed approach. Fig.~\ref{fig: sparse graph + VBound} 
demonstrates that, for systems that are power-balanced but contain under-voltage buses, our proposed method successfully restores voltages within safe bounds by applying compensations at only a few identified nodes. We then examine the 2383-bus system (which remains feasible under a load factor of 1 but exhibits 38 under-voltage buses) to gain further insights into the voltage magnitudes. Fig.~\ref{fig: voltage change}(top) plots the voltage magnitude of each bus returned by the two methods, while Fig. \ref{fig: voltage change}(bottom) focuses on the voltage-violated buses, whose voltage magnitudes are restored back within a safe range by the proposed method. Finally, Fig. \ref{fig: feasibility histogram} shows a histogram of magnitude of slack sources responsible for the voltage restoration. 

Fig. \ref{fig: infeasible, sparse graph + VBound} further evaluates systems experiencing both blackout-induced collapses and voltage violations, demonstrating that our proposed method can simultaneously restore power balance and maintain voltage stability.

\subsection{Scalability}
\begin{figure}[h]
    \centering
    \includegraphics[width=\linewidth]{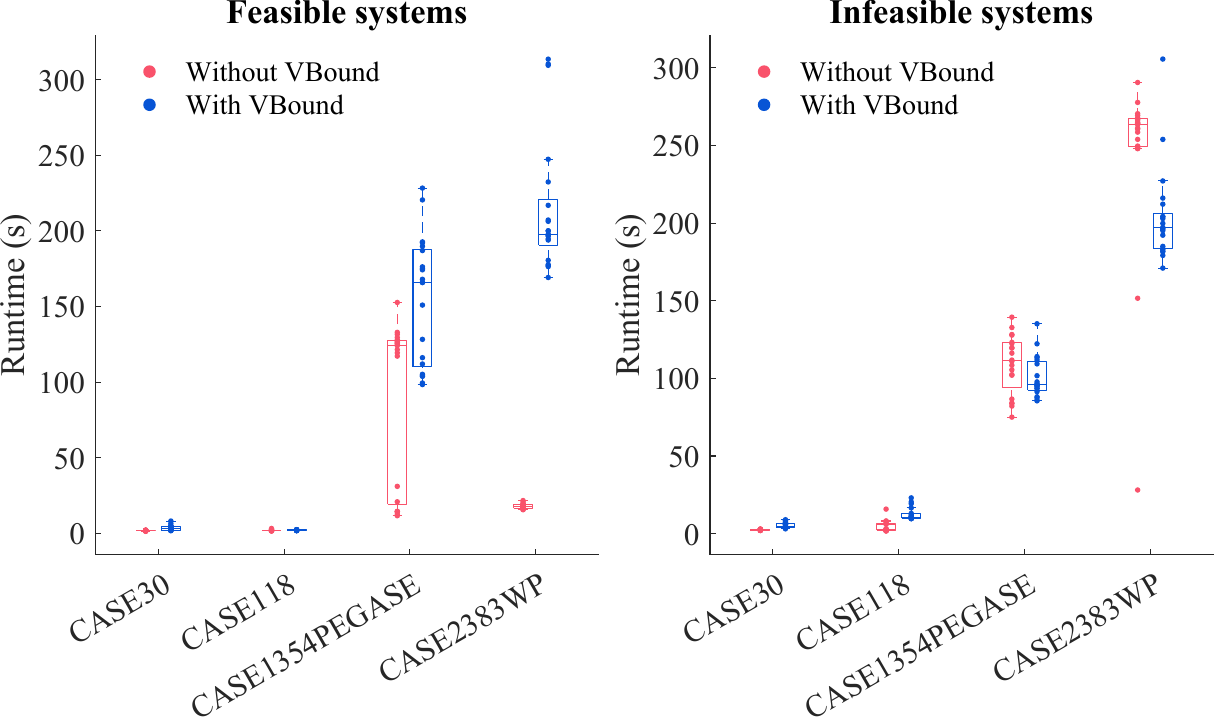}
    \caption{Runtime and scalability. \textit{Left} plot tests on high demand conditions (load factors: 1.4$\sim$1.6 for \textsc{case30}, 1.05$\sim$1.25 for \textsc{case118}, 1.25$\sim$1.45 for \textsc{case1354pegase}, and 1.00$\sim$1.20 for \textsc{case2383wp}) where systems remain feasible (power balanced) but have voltage violations. Under these conditions, the baseline method quickly returns zero compensations, but the proposed method needs longer time because it needs extra loops to  return the sparse compensations for fixing voltage. \textit{Right} plot further increases demand (load factors: 4.20$\sim$4.40 for \textsc{case30}, 1.40$\sim$1.60 for \textsc{case118}, 2.00$\sim$2.20 for \textsc{case1354pegase}, and 1.34$\sim$1.54 for \textsc{case2383wp}) to make systems infeasible (blackout collapsed) and voltage-violated. Under such conditions, our proposed method runs faster than the baseline, as it restores both power balance and voltage profile, resulting in slightly denser compensations than restoring balance alone, and therefore requiring fewer sparsity-enforcing iterations. This overall indicates that actionable compensations are slightly denser yet converge faster.}
    \label{fig: runtime}
\end{figure}

To assess scalability, we evaluate the algorithm’s runtime (in seconds) across systems of varying sizes and load factors. Fig. \ref{fig: runtime} shows that the method scales well to large-scale systems, taking averagely $<4$ min on 2000+ bus systems, and even achieving faster speed than the baseline under blackout conditions.  

\begin{figure*}[h]
\centering
\subfloat[]
{%
  \includegraphics[clip,width=0.33\linewidth]{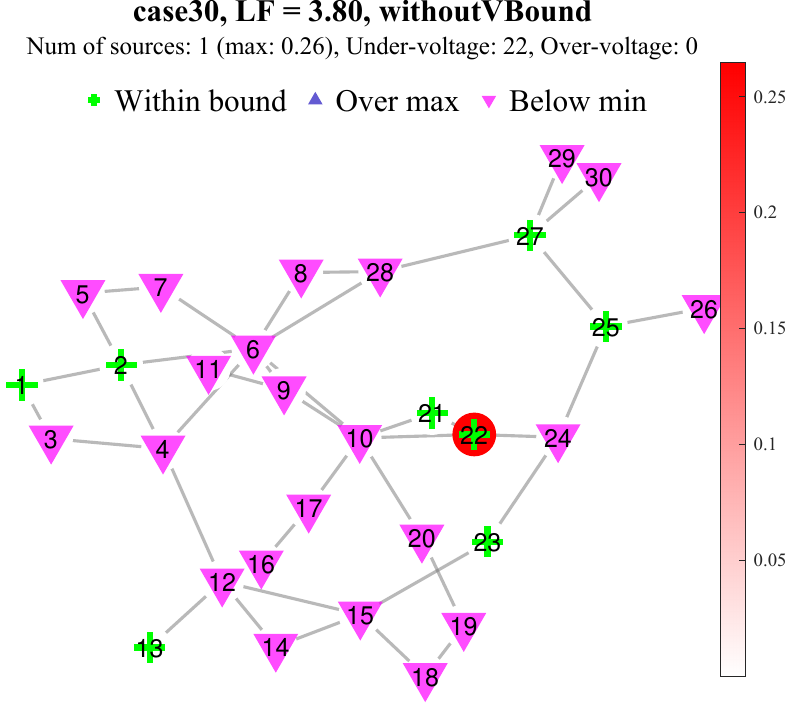}%
}\hfill
\subfloat[]{%
  \includegraphics[clip,width=0.33\linewidth]{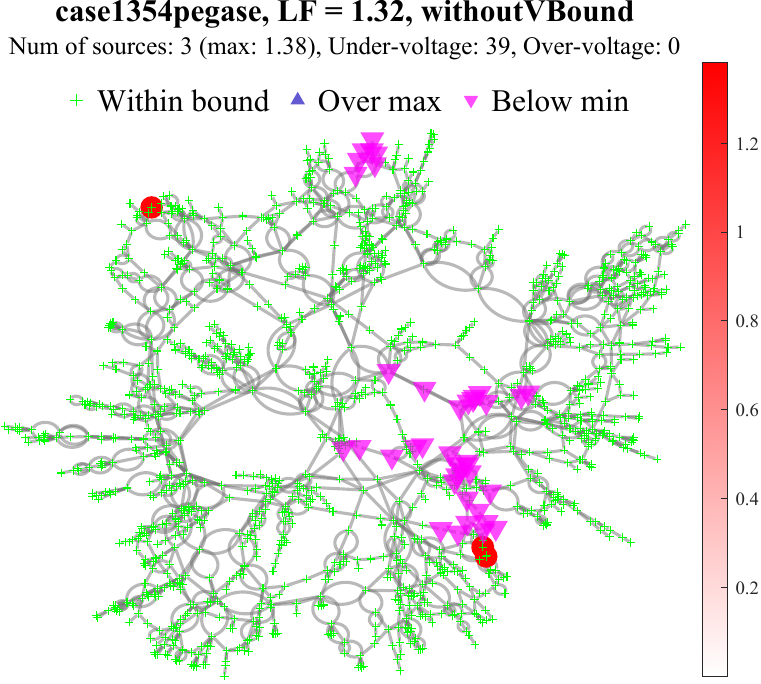}%
}\hfill
\subfloat[]{%
  \includegraphics[clip,width=0.33\linewidth]{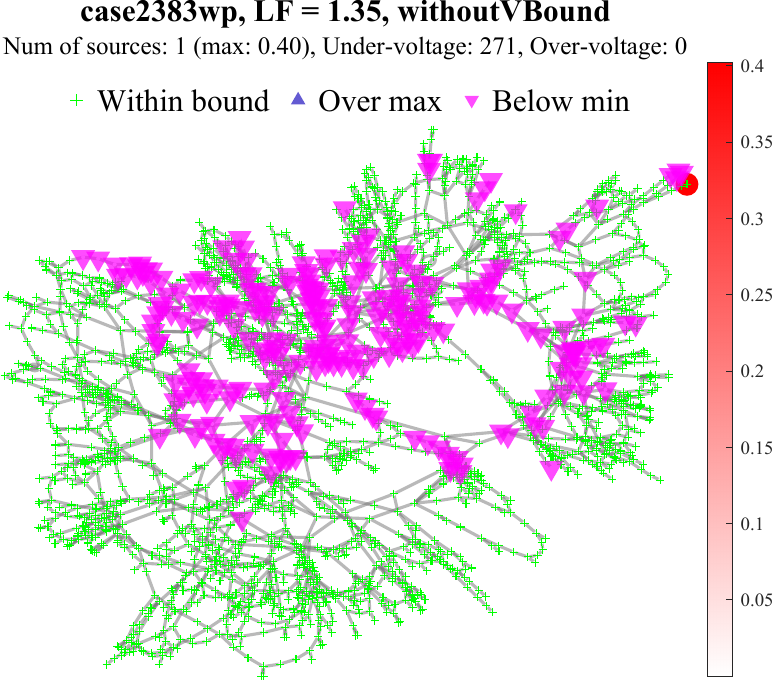}%
}\\
\vspace{5pt}
\subfloat[]{%
  \includegraphics[clip,width=0.33\linewidth]{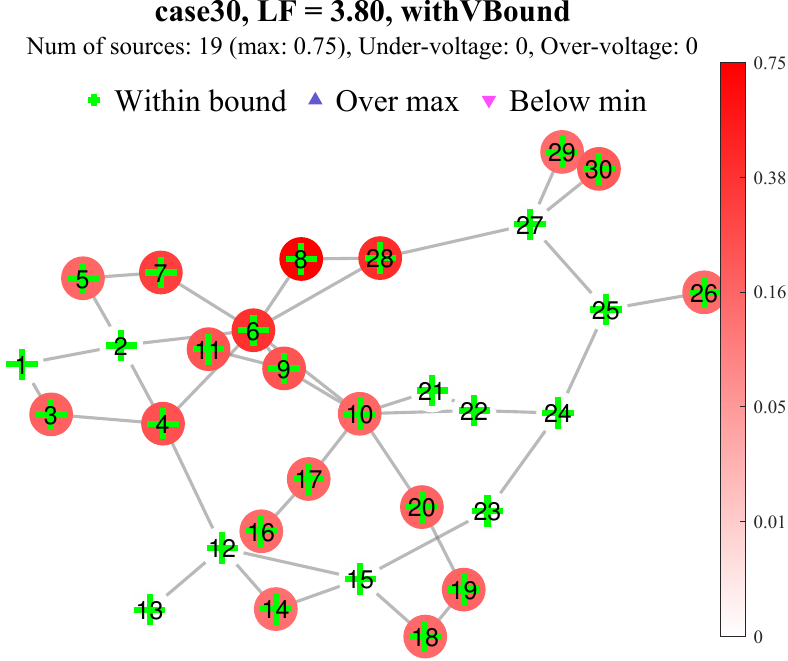}%
}\hfill
\subfloat[]{%
  \includegraphics[clip,width=0.33\linewidth]{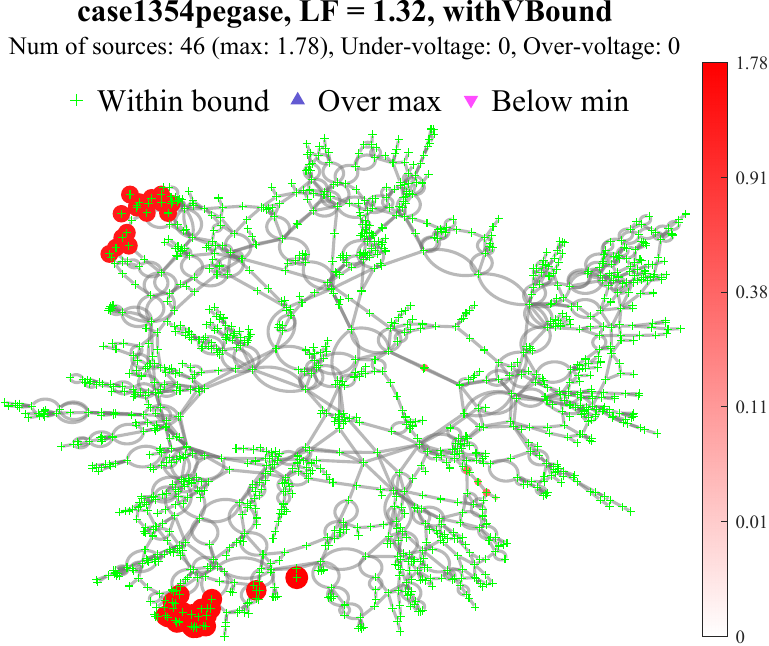}%
}\hfill
\subfloat[]{%
  \includegraphics[clip,width=0.33\linewidth]{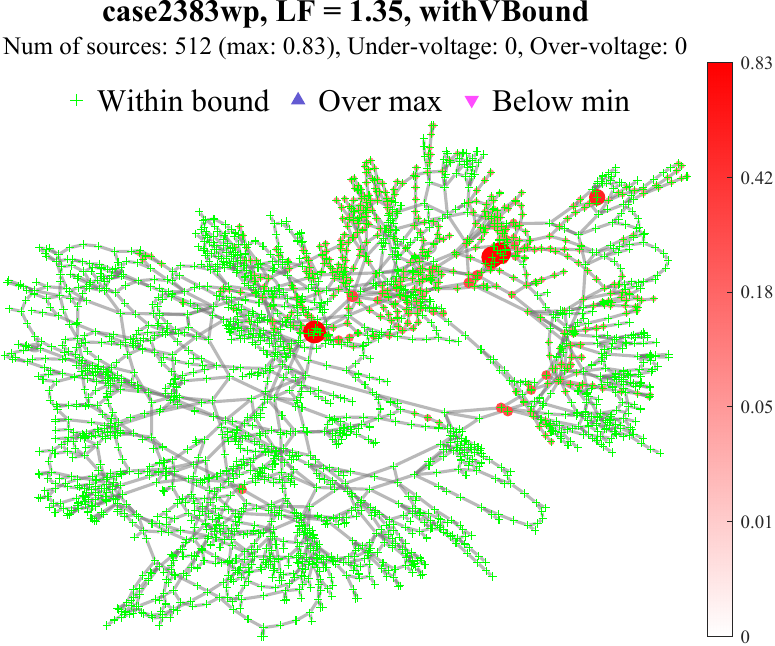}%
}
\caption{High demand makes systems collapse (infeasible). \textit{Top Row (baseline)}: The prior work of sparse diagnosis restores only power balance by compensating at a few identified locations. \textit{Bottom Row (proposed)}: Our proposed approach identifies sparse compensations that restore both power balance and voltage stability, yielding more actionable solutions.}
\label{fig: infeasible, sparse graph + VBound}
\end{figure*}

\section{Conclusion}
\label{sec:Conclusion}
In this paper, we proposed a novel voltage-regulated sparse optimization to identify the minimum set of vulnerable locations responsible for blackouts and voltage instability. When voltage violates the operational bounds, our proposed method yields localized compensation at only a few identified buses to restore voltage stability and, if the system is collapsed, to restore power balance. We conducted experiments on IEEE transmission systems ranging from 30-bus to 2383-bus networks. An instance of the results showed that the 1354-bus system suffers under-voltage at 5 locations when demand increases by 25\%, and the voltage collapse can be fixed by compensating at 20 locations among the 1354 buses in total. The evaluation of runtime also validated the scalability of our method to large-scale systems. 

Findings of this work can serve as a backbone for a more comprehensive and actionable decision-making engine. Additional operational constraints, such as transmission line thermal limits, can be incorporated in a similar way. Moreover, concrete models of compensation resources, such as FACTS devices, battery storage units, and other controllable assets, can be also integrated into this framework.

\balance

\bibliographystyle{IEEEtran}
\bibliography{refbib}

\end{document}